\documentclass{appolb}
\usepackage{epsfig}

\begin{document}
\title{Standard Cosmology in the DGP Brane
 Model
}
\author{Rainer Dick
\address{Department of Physics and Engineering Physics,
University of Saskatchewan,\\ 116 Science Place,
Saskatoon, SK S7N 5E2, Canada}
}
\maketitle
\begin{abstract}
Large extra dimensions provide interesting extensions of our 
parameter space for gravitational theories.
There exist now brane models which can perfectly
reproduce standard four-dimensional Friedmann cosmology.
These models are not motivated by observations,
but they can be helpful in developing new approaches 
to the dimensionality problem in string theory.

I describe the embedding of standard Friedmann cosmology
in the DGP model, and in particular the realization of our current
(dust $+$ $\Lambda$)-dominated universe in this model.
\end{abstract}
\PACS{04.50.+h, 98.80.Cq, 98.80.Hw}
  
\section{Introduction}

In recent years large extra dimensions which can only be probed
by gravitons and eventually non-standard matter have attracted
a lot of attention. These models usually yield the correct Newtonian
 $(1/r)$-potential at large distances because the gravitational
field is quenched on submillimeter transverse scales. This quenching
appears either
due to finite extension of the transverse dimensions \cite{ADD1,AADD}
or due to submillimeter transverse curvature scales
induced by negative cosmological constants\footnote{Please
consult e.g. \cite{rd1,stefan} for much more extensive lists of references.}
 \cite{RS,MVV,GRS,AIMVV,cvetic,banm}.
A common feature of both of these types of models and also
of the old Kaluza--Klein type models is the prediction
of deviations from four-dimensional Einstein gravity at short
distances. If the transverse
length scale is not too small this
implies the possibility to generate bulk gravitons 
in accelerators \cite{AADD,peskin,ACD,rizzo} or stars
\cite{cullen,barger,cassisi,marek,georg}.

In this regard the
recent model of Dvali, Gabadadze and Porrati (DGP) \cite{DGP}
(see also \cite{DG,DGKN} for extensions)
is very different: It predicts that four-dimensional 
Einstein gravity is a short-distance phenomenon 
with deviations showing up at large distances.
The transition between four- and higher-dimensional
gravitational potentials 
in the DGP model arises as a consequence of the presence of both brane and
bulk Einstein terms in the action.

 Furthermore, it was observed in \cite{rd1}
that the DGP model allows for an embedding of standard Friedmann
cosmology in the sense that
the cosmological evolution of the background metric on the brane
can entirely be described by the standard Friedmann equation plus
energy conservation on the brane.
This was later generalized to arbitrary number of transverse dimensions
in \cite{CV}. 

In Sec. \ref{cosmology} I review the standard 
embedding of Friedmann cosmology found in \cite{rd1}, and describe in
particular the realization of a (dust $+$ $\Lambda$)-dominated universe 
in this framework.

\section{The DGP model}\label{dgp}

The action of the DGP model reads
\begin{equation}\label{actionDGP}
S=
\frac{m_4^3}{2}\int dt\int d^{3}\vec{x}
\int dx^\perp
\sqrt{-g}R
\end{equation}
\[
+\left.\int dt\int d^{3}\vec{x}\,\left(
\frac{m_3^2}{2}\sqrt{-g}R^{(d-1)}
-m_4^3\sqrt{-g}\,
\overline{K}
+\mathcal{L}\right)\right|_{x^\perp=0},
\]
where Gaussian normal coordinates 
are employed:
\begin{equation}\label{metric}
ds^2=g_{\mu\nu}dx^\mu dx^\nu+(dx^{\perp})^2.
\end{equation}
The transverse coordinate $|x^\perp|$ is a distance along
orthogonal geodesics to the brane.

The $(3+1)$-dimensional submanifold $x^\perp=0$ is usually 
denoted as a 3-brane, and
$\mathcal{L}$ contains the matter
degrees of freedom on this brane. 
Extrinsic curvature effects have been taken
into account through a Gibbons--Hawking term
 \cite{GH,HL,CR,rdplb2,lalak} (which requires averaging over
the two sides of the brane \cite{rd1}), and
 $m_4$ and $m_3$ are reduced 
Planck masses in five and four dimensions, respectively.

The action (\ref{actionDGP}) yields Einstein equations
\[
m_4^3\left(R_{MN}-\frac{1}{2}g_{MN}R\right)
+m_3^2g_M{}^\mu g_N{}^\nu
\left(R^{(d-1)}_{\mu\nu}
-\frac{1}{2}g_{\mu\nu}R^{(d-1)}\right)\delta(x^\perp)
\]
\begin{equation}\label{einsteinDGP}
=
g_M{}^\mu g_N{}^\nu T_{\mu\nu}\delta(x^\perp),
\end{equation}
corresponding to matching conditions 
\begin{equation}\label{lanczosDGP}
\lim_{\epsilon\to +0}
\left[K_{\mu\nu}\right]_{x^\perp =-\epsilon}^{x^\perp =\epsilon}=
\left.\frac{1}{m_4^3}\left(T_{\mu\nu}-\frac{1}{d-1}
g_{\mu\nu}g^{\alpha\beta}T_{\alpha\beta}\right)\right|_{x^\perp =0}
\end{equation}
\[
-\left.\frac{m_3^2}{m_4^3}\left(R^{(d-1)}_{\mu\nu}
-\frac{1}{2(d-1)}
g_{\mu\nu}g^{\alpha\beta}R^{(d-1)}_{\alpha\beta}
\right)\right|_{x^\perp =0}
\]
 for the extrinsic curvature of the brane.

The use of Gaussian normal coordinates (\ref{metric}) 
implies that we can impose 
 a harmonic gauge condition
only on the longitudinal coordinates $x^\mu$:
\begin{equation}\label{hggauss}
\partial_\alpha h^\alpha{}_\mu+\partial_\perp h_{\perp\mu}
=\frac{1}{2}\partial_\mu\left(h^\alpha{}_\alpha
+h_{\perp\perp}\right),
\end{equation}
but this is sufficient to get a decoupled equation for
the gravitational potential of a static mass
distribution:\\
The transverse equations in the gauge (\ref{hggauss})
\[
R_{\perp\perp}-R^\alpha{}_\alpha=
\frac{1}{2}\partial_\alpha\partial^\alpha
\left(h^\beta{}_\beta
-h_{\perp\perp}\right)
+\partial_\perp\partial_\alpha h^{\alpha}{}_\perp=0,
\]
\[
R_{\perp\mu}=\frac{1}{2}\left(
\partial_\mu\partial_\alpha h^{\alpha}{}_\perp
-\partial_K\partial^K h_{\perp\mu}\right)
+\frac{1}{4}\partial_\mu\partial_\perp
\left(h_{\perp\perp}-h^\alpha{}_\alpha\right)=0
\]
can be solved by $h_{\perp\mu}=0$, $h_{\perp\perp}=h^\alpha{}_\alpha$,
and the remaining equations take the form
\[
m_4^3(\partial_\alpha\partial^\alpha+\partial_\perp^2)h_{\mu\nu}
+m_3^2\delta(x^\perp)\left(\partial_\alpha\partial^\alpha
h_{\mu\nu}-\partial_\mu\partial_\nu h^\alpha{}_\alpha\right)
\]
\[
=-2\delta(x^\perp)\left(T_{\mu\nu}
-\frac{1}{d-1}\eta_{\mu\nu}\eta^{\alpha\beta}T_{\alpha\beta}\right).
\]
This yields the equation for the gravitational
potential of a 
mass density $\varrho(\vec{r})=M\delta(\vec{r})$ on $\mathcal{M}_{3,1}$:
\begin{equation}\label{UDGP}
m_4^3(\Delta+\partial_\perp^2)U(\vec{r},x^\perp)
+m_3^2\delta(x^\perp)\Delta U(\vec{r},x^\perp)
=\frac{2}{3}M\delta(\vec{r})\delta(x^\perp).
\end{equation}

The resulting potential on the brane is \cite{DGP,rd1}
\begin{equation}\label{Udgp}
U(\vec{r})=-\frac{M}{6\pi m_3^2 r}\left[
\cos\!\left(\frac{2m_4^3}{m_3^2}r\right)-
\frac{2}{\pi}\cos\!\left(\frac{2m_4^3}{m_3^2}r\right)
\mbox{Si}\!\left(\frac{2m_4^3}{m_3^2}r\right)\right.
\end{equation}
\[
\left.
+\frac{2}{\pi}\sin\!\left(\frac{2m_4^3}{m_3^2}r\right)
\mbox{ci}\!\left(\frac{2m_4^3}{m_3^2}r\right)
\right],
\]
with the sine and cosine integrals
\[
\mbox{Si}(x)=\int_0^xd\xi\,\frac{\sin\xi}{\xi},
\]
\[
\mbox{ci}(x)=-\int_x^\infty d\xi\,\frac{\cos\xi}{\xi}.
\]
The DGP model thus predicts a 
transition scale 
\begin{equation}\label{ldgp}
\ell_{DGP}=\frac{m_3^2}{2m_4^3}
\end{equation}
between four-dimensional behavior 
and five-dimensional behavior
of the gravitational potential:
\begin{eqnarray*}
r\ll\ell_{DGP}:\,\,
U(\vec{r})=\!&-&\!\frac{M}{6\pi m_3^2 r}
\left[1+\left(\gamma-\frac{2}{\pi}\right)\frac{r}{\ell_{DGP}}
\right.\\
&+&\!\frac{r}{\ell_{DGP}}\ln\!\left(\frac{r}{\ell_{DGP}}\right)
+\left.\mathcal{O}\!\left(\frac{r^2}{\ell^2_{DGP}}\right)\right],
\\
r\gg\ell_{DGP}:\,\,
U(\vec{r})=\!&-&\!\frac{M}{6\pi^2 m_4^3 r^2}
\left[1-2\frac{\ell^2_{DGP}}{r^2}
+\mathcal{O}\!\left(\frac{\ell^4_{DGP}}{r^4}\right)\right].
\end{eqnarray*}
$\gamma\simeq 0.577$ is Euler's constant.

If we would use the usual value of the reduced Planck mass for $m_3$, 
then
the small $r$ potential would be stronger
than the genuine four-dimensional potential 
by a factor $\frac{4}{3}$ because the coupling of 
the masses on the brane 
to the four-dimensional Ricci tensor is increased
by this factor.
This factor $\frac{4}{3}$ is in agreement with the
tensorial structure of the graviton propagator reported
in \cite{DGP}.
Therefore the four-dimensional reduced Planck mass is slightly larger
in the DGP model than in ordinary Einstein gravity:
\begin{equation}\label{modplanck}
 m_3=(6\pi G_{N,3})^{-1/2}\simeq 2.8\times 10^{18}\,\mathrm{GeV}.
\end{equation}

The potential is displayed in Fig. 1.

\begin{figure}[htb]
\vspace*{35mm}
\begin{center}
\hspace*{35mm}\scalebox{0.47}{\includegraphics{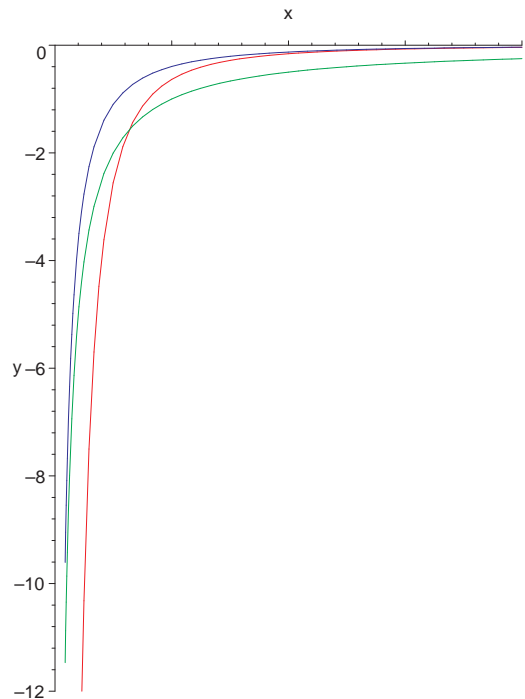}}
\label{fig:dggpot}
\caption{The blue line is the gravitational potential in the DGP model 
as a function
of $x=r/\ell_{DGP}$. 
The horizontal axis covers the region $0\le x\le 4$.
The vertical units correspond to $M/(12m_4^3)$. 
The
green line is the ordinary three-dimensional Newton potential in these 
units, and the red line
is the corresponding potential in four spatial dimensions.}
\end{center}
\end{figure}

The current limit on deviations from Einstein gravity at {\em large
distances} is still set
by \cite{talmadge}, see also \cite{ephraim, will}.

The limit $\ell_{DGP}> 10^{14}\,$m would translate with
(\ref{ldgp},\ref{modplanck}) 
into a bulk Planck mass $m_4<200\,$GeV. This may seem surprisingly low,
but recall from (\ref{Udgp}) that the relevant graviton coupling scale at 
distances well below $\ell_{DGP}$ is the large Planck mass $m_3$ on the brane,
and lower $m_4$ means larger $\ell_{DGP}$, making it even
harder to detect any deviations from
Einstein gravity.

It is certainly easy to constrain $\ell_{DGP}$ to supergalactic scales,
because the DGP model predicts a weakening of gravity at large distances,
thus potentially increasing the need for dark matter.

However, interest in this model does not arise from the hope that one
might detect any corresponding effects at galactic or not too large
supergalactic scales: The interest in the model results from the 
observation that it provides a simple, yet surprising mechanism
to accomodate four-dimensional gravity in a model with infinitely large
extra dimensions. 

\section{Standard cosmology in the DGP model}\label{cosmology}

 From the fact that the DGP model predicts deviations only at large distances one
might hope that it could be ruled out from cosmological observations, but we 
will see that it can account for standard Friedmann cosmology at any
distance scale on the brane:

Brane cosmology usually starts from the line element
(with $x_i\equiv x^i$, $r^2\equiv x_ix^i$)
\begin{equation}\label{cosprinc}
ds^2=-n^2(x^\perp,t)dt^2
\end{equation}
\[
+a^2(x^\perp,t)
\!\left(\delta_{ij}+k\frac{x_i x_j}{1-kr^2}
\right)dx^idx^j
+b^2(x^\perp,t)d{x^\perp}^2.
\]
This {\em ansatz} implies a brane cosmological
principle in that it assumes that every hypersurface
 $x^\perp=$ const.\ is a Robertson--Walker spacetime
with cosmological time $T|_{x^\perp}=\int |n(x^\perp,t)|dt$. 

Building on the results of \cite{BDL,BDEL},
the cosmological evolution equations of a 3-brane in 
a five-dimensional bulk following from
(\ref{einsteinDGP}) and (\ref{lanczosDGP}) were presented in
 \cite{cedric,DDG}.

Here I will follow \cite{rd1} and give the results for a brane
of dimension $\nu+1$.

 The Einstein tensors for the metric (\ref{cosprinc})
in Gaussian normal coordinates ($b^2=1$) and in $d=\nu+1$ 
spatial dimensions are\\
on the hypersurfaces $x^\perp=$ const.:
\begin{equation}\label{gbrane00}
G_{00}^{(\nu)}=\frac{1}{2}\nu(\nu-1)n^2
\!\left(\frac{\dot{a}^2}{n^2a^2}+\frac{k}{a^2}\right)
\end{equation}
\begin{eqnarray}\label{gbraneij}
G_{ij}^{(\nu)}=(\nu-1)\!\left(
\frac{\dot{n}\dot{a}}{n^3a}-\frac{\ddot{a}}{n^2a}\right)g_{ij}
-\frac{1}{2}(\nu-1)(\nu-2)\!\left(
\frac{\dot{a}^2}{n^2a^2}+\frac{k}{a^2}\right)g_{ij},
\end{eqnarray}
and in the bulk:
\begin{equation}\label{bdl00}
G_{00}=\frac{1}{2}\nu(\nu-1)n^2
\!\left(\frac{\dot{a}^2}{n^2a^2}-\frac{{a'}^2}{a^2}
+\frac{k}{a^2}\right)-\nu n^2\frac{a''}{a},
\end{equation}
\begin{eqnarray}\label{bdlij}
G_{ij}&=&\frac{1}{2}(\nu-1)(\nu-2)\!\left(\frac{{a'}^2}{a^2}
-\frac{\dot{a}^2}{n^2a^2}-\frac{k}{a^2}\right)g_{ij}
\\
 &&\!\! +(\nu-1)\!\left(\frac{a''}{a}+\frac{n'a'}{na}
-\frac{\ddot{a}}{n^2a}+\frac{\dot{n}\dot{a}}{n^3a}\right)g_{ij}
+\frac{n''}{n}g_{ij},\nonumber
\end{eqnarray}
\begin{equation}\label{bdl0perp}
G_{0\perp}=\nu\!\left(
\frac{n'}{n}\frac{\dot{a}}{a}-\frac{\dot{a}'}{a}
\right),
\end{equation}
\begin{equation}\label{bdlpp}
G_{\perp\perp}=\frac{1}{2}\nu(\nu-1)
\!\left(\frac{{a'}^2}{a^2}-\frac{\dot{a}^2}{n^2a^2}
-\frac{k}{a^2}\right)
+\nu\!\left(\frac{n'a'}{na}+\frac{\dot{n}\dot{a}}{n^3a}
-\frac{\ddot{a}}{n^2a}\right).
\end{equation}

The matching conditions (\ref{lanczosDGP}) for
an ideal fluid on the brane 
\[
T_{00}=\varrho n^2,\,\, T_{ij}=p g_{ij}
\]
read
\begin{eqnarray}\label{coslanczoseq00}
\lim_{\epsilon\to +0}\left[
\partial_\perp n\right]_{x^\perp =-\epsilon}^{x^\perp =\epsilon}
&=&\frac{n}{\nu m_{\nu+1}^\nu}\left.\!\Bigg(
(\nu-1)\varrho+\nu p\Bigg)\right|_{x^\perp =0}\\
&&\!\! +\frac{m_{\nu}^{\nu-1}}{m_{\nu+1}^\nu}(\nu-1)n\left.\!\left(
\frac{\ddot{a}}{n^2a}-\frac{\dot{a}^2}{2n^2a^2}-
\frac{\dot{n}\dot{a}}{n^3a}-\frac{k}{2a^2}\right)\right|_{x^\perp =0},
\nonumber
\end{eqnarray}
\begin{equation}\label{coslanczoseqij}
\lim_{\epsilon\to +0}\left[
\partial_\perp a\right]_{x^\perp =-\epsilon}^{x^\perp =\epsilon}
=\frac{m_{\nu}^{\nu-1}}{2m_{\nu+1}^\nu}(\nu-1)\left.\!\left(
\frac{\dot{a}^2}{n^2a}+\frac{k}{a}\right)\right|_{x^\perp =0}
-\left.\frac{\varrho a}{\nu m_{\nu+1}^\nu}\right|_{x^\perp =0}.
\end{equation}
This
corresponds to effective gravitational contributions to
the pressure and energy density on the brane:
\[
\varrho_G=-\frac{1}{2}\nu(\nu-1)m_{\nu}^{\nu-1}\!\left(
\frac{\dot{a}^2}{n^2a^2}+\frac{k}{a^2}\right),
\]
\[
p_G=(\nu-1)m_{\nu}^{\nu-1}\!\left(
\frac{\ddot{a}}{n^2a}-\frac{\dot{n}\dot{a}}{n^3a}\right)
+\frac{1}{2}(\nu-1)(\nu-2)m_{\nu}^{\nu-1}\!\left(
\frac{\dot{a}^2}{n^2a^2}+\frac{k}{a^2}\right).
\]

Energy conservation on the brane follows from
the absence of transverse momentum, $T_{0\perp}=0$.
With (\ref{bdl0perp}) this implies
\begin{equation}\label{0perp0}
\frac{n'}{n}=\frac{\dot{a}'}{\dot{a}}
\end{equation}
and in particular
\[
\lim_{\epsilon\to +0}\left[\frac{n'}{n}
\right]_{x^\perp =-\epsilon}^{x^\perp =\epsilon}
=\lim_{\epsilon\to +0}\left[\frac{\dot{a}'}{\dot{a}}
\right]_{x^\perp =-\epsilon}^{x^\perp =\epsilon}.
\]
Insertion of (\ref{coslanczoseq00},\ref{coslanczoseqij})
into this equation yields the sought for conservation
equation
\begin{equation}\label{econ}
\dot{\varrho}a\Big|_{x^\perp=0}=-\nu(\varrho+p)\dot{a}\Big|_{x^\perp=0}.
\end{equation}

Insertion of (\ref{0perp0}) into (\ref{bdl00}) and (\ref{bdlpp})
for $x^\perp\neq 0$
yields a $\nu$-dimensional version of the integral
of Bin\'{e}truy {\it et al}.\ \cite{BDEL}:
\[
\frac{2}{\nu n^2}a'a^\nu G_{00}=
\frac{\partial}{\partial x^\perp}\!\left(
\frac{\dot{a}^2}{n^2}a^{\nu-1}-
{a'}^2a^{\nu-1}+ka^{\nu-1}
\right)=0,
\]
\[
\frac{2}{\nu}\dot{a}a^\nu G_{\perp\perp}=
-\frac{\partial}{\partial t}\!\left(
\frac{\dot{a}^2}{n^2}a^{\nu-1}-
{a'}^2a^{\nu-1}+ka^{\nu-1}
\right)=0,
\]
i.e.
\begin{equation}\label{intbdel+}
I^{+} =\left.
\left(\frac{\dot{a}^2}{n^2}-
{a'}^2+k\right)a^{\nu-1}\right|_{x^\perp >0}
\end{equation}
and
\begin{equation}\label{intbdel-}
I^{-} =\left.
\left(\frac{\dot{a}^2}{n^2}-
{a'}^2+k\right)a^{\nu-1}\right|_{x^\perp <0}
\end{equation}
are two constants, with $I^{+} =I^{-} $
if 
\[
\lim_{\epsilon\to +0}a'\Big|_{x^\perp=\epsilon}
=\pm\lim_{\epsilon\to +0}a'\Big|_{x^\perp=-\epsilon}.
\]

We have not yet taken into account $G_{ij}=0$ in the bulk.
However, Eq.\ (\ref{0perp0}) implies 
$\partial_\perp(n/\dot{a})=0$, and therefore
\[
\frac{n''}{n}=\frac{\dot{a}''}{\dot{a}}.
\]
This, the bulk equations $G_{00}=G_{\perp\perp}=0$, and the constancy
of $I^{\pm}$ imply that the bulk equation $G_{ij}=0$
is already satisfied and does not provide any new information.

We can now simplify the previous equations by further restricting
our Gaussian normal coordinates through the gauge 
\begin{equation}\label{frwgauge}
n(0,t)=1
\end{equation}
by simply performing the transformation
\[
t\,\,\Rightarrow\,\,t_{FRW}=\int^tdt'\,n(0,t')
\]
of the time coordinate.
This gauge is convenient because it gives the usual
cosmological time on the brane. Henceforth this gauge will be
adopted, but the index FRW will be omitted.

Eqs.\ (\ref{0perp0},\ref{frwgauge}) imply that our 
basic dynamical variable
is $a(x^\perp,t)$, with $n(x^\perp,t)$ given by
\[
n(x^\perp,t)=\frac{\dot{a}(x^\perp,t)}{\dot{a}(0,t)}.
\]
The basic set of cosmological equations in the present setting 
(without a cosmological
constant in the bulk) are thus
eqs.\ (\ref{coslanczoseqij},\ref{econ},\ref{intbdel+},\ref{intbdel-}),
 which have to be amended
with dispersion relations (or corresponding evolution equations) for the 
ideal fluid
components on the brane:\\[2ex]
\fbox{
\begin{minipage}{0.97\textwidth}
\vspace*{2ex}
\[
\lim_{\epsilon\to +0}\left[
\partial_\perp a\right]_{x^\perp =-\epsilon}^{x^\perp =\epsilon}(t)
=\frac{m_{\nu}^{\nu-1}}{2m_{\nu+1}^\nu}(\nu-1)
\frac{\dot{a}^2(0,t)+k}{a(0,t)}
-\frac{\varrho(t) a(0,t)}{\nu m_{\nu+1}^\nu},
\]
\[
I^{+} =\left.
\left(\dot{a}^2(0,t)-
{a'}^2(x^\perp,t)+k\right)a^{\nu-1}(x^\perp,t)\right|_{x^\perp > 0},
\]
\[
I^{-} =\left.
\left(\dot{a}^2(0,t)-
{a'}^2(x^\perp,t)+k\right)a^{\nu-1}(x^\perp,t)\right|_{x^\perp < 0},
\]
\[
\dot{\varrho}(t)a(0,t)=-\nu(\varrho(t)+p(t))\dot{a}(0,t),
\]
\[
p(t)=p(\varrho(t)),
\]
\[
n(x^\perp,t)=\frac{\dot{a}(x^\perp,t)}{\dot{a}(0,t)}.
\]
\vspace*{1ex}
\end{minipage}
}

\vspace*{2ex}
Our primary concern with regard to observational consequences is the
evolution of the scale factor $a(0,t)$ on the brane, and we can use
the integrals $I^{\pm}$ to eliminate
the normal derivatives $a'(x^\perp\to\pm 0,t)$ from the brane analogue 
of the Friedmann equation:
\[
\pm\sqrt{\dot{a}^2(0,t)+k-I^+ a^{1-\nu}(0,t)}
\mp\sqrt{\dot{a}^2(0,t)+k-I^- a^{1-\nu}(0,t)}
\]
\begin{equation}\label{friedmann1}
=\frac{m_{\nu}^{\nu-1}}{2m_{\nu+1}^\nu}(\nu-1)
\frac{\dot{a}^2(0,t)+k}{a(0,t)}
-\frac{\varrho(t) a(0,t)}{\nu m_{\nu+1}^\nu}.
\end{equation}
If this equation is solved for $a(0,t)$
by using the dispersion relation and energy
conservation on the brane, then $a(x^\perp,t)$ can be determined in the
bulk from the constancy of $I^{\pm}$.

There must be at least one minus sign on the left hand side of 
(\ref{friedmann1}) if the right hand side is negative, but the dynamics
of the problem does not require symmetry across the brane.
The constants $I^{\pm}$ must be considered as initial conditions,
and if e.g.\
 $I^+\neq I^-$, then there cannot be any symmetry across the brane.

If $m_\nu\neq 0$ and the 
normal derivatives on the
brane have the same sign:
\begin{equation}\label{ordincond}
m_\nu a'(x^\perp\to +0,t)a'(x^\perp\to -0,t)>0,
\end{equation}
 then the cosmology of our brane approximates
ordinary Friedmann--Robertson--Walker cosmology during those epochs
when
\[
I^{\pm} \ll \left(\dot{a}^2(0,t)+k\right)a^{\nu-1}(0,t).
\]
In particular, this applies to late epochs in expanding open or flat
branes ($k\neq 1$).

\subsection{The embedding of standard Friedmann cosmology}
\label{sec:general}

Standard cosmology may be realized in the DGP model
in an even more direct way:

If (\ref{ordincond}) holds
and $I^+=I^-$, then (\ref{friedmann1}) reduces {\em entirely} to 
the ordinary Friedmann equation for a $(\nu+1)$-dimensional
spacetime \cite{rd1}. This embedding of standard Friedmann cosmology is then
given by the following set of cosmological evolution equations:
\\[2ex]
\fbox{
\begin{minipage}{0.97\textwidth}
\vspace*{2ex}
\[
\frac{\dot{a}^2(0,t)+k}{a^2(0,t)}
=\frac{2\varrho(t)}{\nu(\nu-1) m_{\nu}^{\nu-1}},
\]
\[
I=
\left(\dot{a}^2(0,t)-
{a'}^2(x^\perp,t)+k\right)a^{\nu-1}(x^\perp,t),
\]
\[
\dot{\varrho}(t)a(0,t)=-\nu(\varrho(t)+p(t))\dot{a}(0,t),
\]
\[
p(t)=p(\varrho(t)),
\]
\[
n(x^\perp,t)=\frac{\dot{a}(x^\perp,t)}{\dot{a}(0,t)}.
\]
\vspace*{1ex}
\end{minipage}
}

\vspace*{2ex}

The evolution of the background geometry of the observable
universe according to the Friedmann equation can thus be embedded
 in the DGP model, with the bevavior 
of $a(x^\perp,t)$ off the brane determined solely by the integral 
 $I$ and the boundary condition $a(0,t)$ from the Friedmann
equation. 

This embedding will be asymmetric in all realistic cases, because
the requirement that the Friedmann equation holds on the brane is equivalent
to the smoothness condition
\[
\lim_{\epsilon\to +0}
\partial_\perp a(\epsilon,t)=
\lim_{\epsilon\to +0}
\partial_\perp a(-\epsilon,t).
\]
This could yield a symmetric embedding only for $a'(0,t)=0$, but this
is incompatible with the time-independence of the integral $I$
(apart from the exotic case $k=-1$, $\dot{a}^2=1$). 
And {\it vice versa}: The previously often employed 
assumption that embeddings would have to
be symmetric implied a cusp at the brane and a corresponding violation
of the Friedmann equation on the brane. 
Observation of the Hubble flow thus might have had observable
consequences on brane cosmology, see \cite{AM,DDG2} for a discussion
of this, but in the present
embedding scenarios the Hubble flow cannot rule out brane
scenarios.

In the sequel we will choose the sign of $x^\perp$ in the direction of
increasing scale factor: 
\begin{equation}\label{eq:sign}
a'(0,t)>0.
\end{equation}

The possibility of a direct embedding of Friedmann cosmology 
is a consequence of the fact that the evolution of the background
geometry (\ref{cosprinc}) and the source terms $\varrho$, $p$
are supposed to depend only on $t$ and $x^\perp$.
This implies the possibility to decouple the brane and the bulk contributions
in the Einstein equation for the background metric, and
in this case deviations
from Friedmann--Robertson--Walker cosmology would only show up
in specific $\vec{x}$-dependent effects like the evolution of
cosmological perturbations and structure formation.

In the relevant case $\nu=3$ we find from the second equation
\[
I=
\left(\dot{a}^2(0,t)-
{a'}^2(x^\perp,t)+k\right)a^{2}(x^\perp,t),
\]
and from the equation for $n(x^\perp,t)$, the solutions for the metric
components off the brane in terms of the metric on the brane
(with the sign convention from (\ref{eq:sign})):
\begin{equation}\label{eq:axt}
a^2(x^\perp,t)=a^2(0,t)+\left(\dot{a}^2(0,t)+k\right){x^\perp}^2
 + 2\sqrt{\left(\dot{a}^2(0,t)+k\right)
a^2(0,t)-I}x^\perp,
\end{equation}
\vspace*{0.3ex}
\[
n(x^\perp,t)=\left[a(0,t)+\ddot{a}(0,t){x^\perp}^2
+
a(0,t)x^\perp\frac{a(0,t)\ddot{a}(0,t)+\dot{a}^2(0,t)+k}{
\sqrt{\left(\dot{a}^2(0,t)+k\right)a^2(0,t)-I}}\right]
\]
\begin{equation}\label{eq:nxt}
\times
\left[a^2(0,t)+\left(\dot{a}^2(0,t)+k\right){x^\perp}^2
 + 2\sqrt{\left(\dot{a}^2(0,t)+k\right)
a^2(0,t)-I}x^\perp
\right]^{-1/2}.
\end{equation}

This embedding of Friedmann cosmology on the brane becomes
particularly simple for $I=0$:
\begin{equation}\label{eq:axtI=0}
a(x^\perp,t)=a(0,t)+\sqrt{\dot{a}^2(0,t)+k}x^\perp,
\end{equation}
\begin{equation}\label{eq:nxtI=0}
n(x^\perp,t)=1+\frac{\ddot{a}(0,t)}{\sqrt{\dot{a}^2(0,t)+k}}x^\perp.
\end{equation}

\subsection{Radiation dominated spatially flat universe in the DGP
model}

 For early universe cosmology spatial curvature and the
cosmological constant are negligible compared to the radiation dominated
matter density.

The energy density and scale factor on the brane evolve in the standard way
\[
\varrho(t)=\frac{3m_3^2}{4t^2},
\]
\begin{equation}\label{initcond}
a(0,t)=C\sqrt{t},
\end{equation}
and the metric off the brane is given by \cite{rd1}
\[
a^2(x^\perp,t)=\frac{C^2}{4t}{x^\perp}^2+\sqrt{C^4-4I}x^\perp
+C^2t
\]
and
\[
n^2(x^\perp,t)=\frac{C^2}{4t^2}
\frac{\left(4t^2-{x^\perp}^2\right)^2}{C^2{x^\perp}^2
+4\sqrt{C^4-4I}x^\perp t+4C^2t^2}.
\]
This yields in particular for $I=0$:
\[
a(x^\perp,t)=C\!\left(\sqrt{t}+\frac{x^\perp}{2\sqrt{t}}\right),
\]
\[
n(x^\perp,t)=1-\frac{x^\perp}{2t}.
\]

There appear coordinate singularities on the spacelike hypercone
 $x^\perp=\pm 2t$. This is presumably a consequence of the fact that 
the orthogonal geodesics emerging from the brane (which we used to set
up our Gaussian normal system) do not cover the full five-dimensional
spacetime.

\subsection{(Dust $+$ $\Lambda$)-dominated universe in the DGP model}

The evolution of the background metric in a (dust $+$ $\Lambda$)-dominated 
universe is readily inferred from
energy conservation and the Friedmann equation. The energy density in the dust
evolves according to 

\begin{equation}\label{eq:rho2}
\varrho_{dust}=\frac{\Lambda}{\sinh^2\!\left(\frac{\sqrt{3\Lambda}}{2m_3}t\right)},
\end{equation}
and the scale factor on the brane is
\begin{equation}\label{eq:a0t2}
a(0,t)=\left(
\frac{\sinh(\sqrt{3\Lambda}t/2m_3)}{\sinh(\sqrt{3\Lambda}t_0/2m_3)}
\right)^{\frac{2}{3}}.
\end{equation}

 From subsection \ref{sec:general} it is clear that minimal embeddings
correspond to $I=0$, and Eqs. (\ref{eq:axtI=0},\ref{eq:nxtI=0}) yield
\begin{equation}\label{eq:axt2}
a(x^\perp,t)=\left(
\frac{\sinh(\sqrt{3\Lambda}t/2m_3)}{\sinh(\sqrt{3\Lambda}t_0/2m_3)}
\right)^{\frac{2}{3}}
\left[
1+\sqrt{\frac{\Lambda}{3}}\coth\!\left(\frac{\sqrt{3\Lambda}}{2m_3}t\right)
\frac{x^\perp}{m_3}
\right],
\end{equation}
\begin{equation}\label{eq:axt2}
n(x^\perp,t)=1+\frac{\ddot{a}(0,t)}{\dot{a}(0,t)}x^\perp
\end{equation}
\[
=1+\frac{\sqrt{3\Lambda}}{2m_3}\left[
\tanh\!\left(\frac{\sqrt{3\Lambda}}{2m_3}t\right)-\frac{2}{3}
\coth\!\left(\frac{\sqrt{3\Lambda}}{2m_3}t\right)
\right]x^\perp.
\]

This solution has to be smoothly connected to the corresponding radiation
dominated solution if aspects of the Hubble flow at matter--radiation
equality are examined. However, for later times the corresponding shift
in $t$ is negligible.

\section{Conclusions}

Brane models provide a somewhat exotic, yet interesting extension of
our parameter space for gravitational theories. We have seen that
even standard Friedmann cosmology on the brane can be accomodated
by brane models with Einstein terms both on the brane and in the bulk,
thus implying that cosmological tests of these models should come from
structure formation, where the DGP model should have problems due to the
weakening of gravity at large distances.

Another matter of concern for brane models in general are gauge fields.
While it may be mathematically possible to restrict non-gravitational
terms in the action {\it a priori} to brane contributions, logically this
may not seem entirely satisfactory. It is relatively easy to
limit the penetration depth for scalars and fermions, but I am not aware
of any satisfactory mechanism to constrain the penetration depth of 
time-dependent gauge fields (static sources are no problem, see \cite{DGS}).

Still, it is of interest to study the properties of these models:
Brane models like the DGP model provide a framework for extensions
of the brane models of string theory to infinitely large transverse
dimensions, thus potentially shedding new light on the dimensionality
problem in string theory. \\[2ex]
\noindent
{\bf Acknowledgement:} I would like to thank Marek Biesiada and the organizers
of the XXV International School of Theoretical Physics for discussions and 
their very kind hospitality. This work was supported in part by NSERC Canada.

\end{document}